# Artificial Neural Network for Cybersecurity: A Comprehensive Review

**Prajoy Podder [1], Subrato Bharati [2], M. Rubaiyat Hossain Mondal [3], Pinto Kumar Paul [4], Utku Kose [5]**

[1,2,3] Institute of Information and Communication Technology, Bangladesh University of Engineering and Technology, Dhaka-1205, Bangladesh
[4]Ranada Prasad Shaha University, Narayanganj, Bangladesh
[5]Suleyman Demirel University, Isparta, Turkey

[1] *prajoypodder@gmail.com*, [2] *subratobharati1@gmail.com*, [3] *rubaiyat97@iict.buet.ac.bd*, [4] *pinto.kumar07@gmail.com*,
[5]*utkukose@sdu.edu.tr*

*Abstract*: Cybersecurity is a very emerging field that protects systems, networks, and data from digital attacks. With the increase in the scale of the Internet and the evolution of cyber attacks, developing novel cybersecurity tools has become important, particularly for Internet of things (IoT) networks. This paper provides a systematic review of the application of deep learning (DL) approaches for cybersecurity. This paper provides a short description of DL methods which is used in cybersecurity, including deep belief networks, generative adversarial networks, recurrent neural networks, and others. Next, we illustrate the differences between shallow learning and DL. Moreover, a discussion is provided on the currently prevailing cyber-attacks in IoT and other networks, and the effectiveness of DL methods to manage these attacks. Besides, this paper describes studies that highlight the DL technique, cybersecurity applications, and the source of datasets. Next, a discussion is provided on the feasibility of DL systems for malware detection and classification, intrusion detection, and other frequent cyber-attacks, including identifying file type, spam, and network traffic. Our review indicates that high classification accuracy of 99.72% is obtained by restricted Boltzmann machine (RBM) when applied to a custom dataset, while long short-term memory (LSTM) achieves an accuracy of 99.80% for KDD Cup 99 dataset. Finally, this article discusses the importance of cybersecurity for reliable and practicable IoT-driven healthcare systems.

## I. Introduction

Cybersecurity is the complete package of all techniques and technologies responsible for defending networks, software, and data from attacks [1, 2]. The mechanism of cyber defense is available at the network, data level, host and application. Some cybersecurity tools like firewalls, the system of intrusion detection, the system of intrusion protection etc., are always active at each end to identify security breaches and stop attacks [3, 4]. Nevertheless, with the increasing number of systems having Internet-connection, the risk of attacks is increasing day by day. With the realization of Internet of things (IoT) networks, cybersecurity is becoming more important than ever. Computer networks including IoT are vulnerable to many security threats. Some attacks are of known pattern can be easily managed. However, attackers are developing zero-day exploits, where the attack takes places as soon as a weakness in the system is detected. Such an attack has no previous record and the attack can damange the computer system before the problem is solved. Moreover, the system must be defended not only from external threats but also need to be protected from insider threats, such as misuse of the authorized access, which can be an individual or mean to be a part of the organization.

The main challenge is finding out the compromising system's indicators from the attack's lifecycle, which may have meaningful signs of a future attack. However, this could be a difficult job because of massive quantities of data-generating continuously from lots of cyber-enabled devices. Data Science uses the extensive range of data made by the cyber defense system, including the security information and event management (SIEM) scheme, sometimes overflowing the specialist in security with the event warnings, identifying patterns, related events, and detecting abnormal behaviour to improve cybersecurity.

Hybrid detection in security amalgamates anomaly and misuse detection. This system is mainly used to decrease the rate of false-positive value of anonymous attacks and enhance the rate of detection of recognized intrusions. Maximum DL approaches are hybrid methods [5, 6].Previous reviews, i.e., those in [7-9] have illustrated applications of machine learning (ML) for the solution of cyber-related problems, but deep learning (DL) methods have not been focused on those papers. Some works illustrate DL approaches for cybersecurity. These approaches have some limitations in the applications on cybersecurity [10, 11].

This paper reviews cybersecurity using DL. Moreover, DL methods in cybersecurity and the difference between DL and shallow learning are broadly discussed, and the results of different DL methods are reported. The rest of the paper is organized as follows. Section II discusses the differences between DL and ML, Section III introduces different DL methods in the context of cybersecurity. DL and shallow learning are compared in Section IV. The performance results of different DL methods are reported in Section V. Finally, the paper concludes in Section VI.

## II. DL AND ML

Both ML and DL are subsets of artificial intelligence (AI). The differences between ML and DL include the following:

a) Dependencies of data: The performance of DL models are not comparatively better than traditional ML models for small-scale data volumes. The reason behind this is DL models need a large portion of data to comprehend the data flawlessly. On the other hand, traditional ML algorithms use the established rules [14].

b) Hardware dependencies: Graphics Processing Unit (GPU) can be considered essential hardware for training the DL models properly. The GPU is mainly applied to optimize matrix processes effectively since DL models require a lot of matrix operations. On the other hand, traditional ML algorithms do not usually require high-performance machines with GPUs [18].

c) Processing in feature: The procedure of driving domain knowledge into a feature extractor in order to decrease the complexity of data is termed feature processing. Patterns are usually generated in feature processing, and therefore, ML and DL algorithms work better. However, this stage is time-consuming, and specialized knowledge is required in this case. The performance of most ML models rely on the features accuracy (i.e., pixel values, textures, shapes, locations, etc.) extracted. Attempting to derive high-level features openly from personal data is a main difference between traditional ML and DL algorithms [17]. Accordingly, DL decreases the designing effort to an extracting features for every problem.

d) Execution time: Large execution time is needed to train a DL model owing to its having various parameters. The training step also takes longer. On the contrary, less execution time (only seconds to few hours) is needed to train a ML model. Nevertheless, the time required in testing stage is just the contrast. DL models need very short testing time compared with some ML models.

## III. DL APPROACHES IN CYBER SECURITY

This section illustrates different types of DL methods used in cyber security.

### A. Deep Belief Networks

Deep Belief Networks (DBNs) is brought in a seminal paper by Geoffrey Hinton. DBNs are a class of Deep Neural Networks (DNNs). A DBN is composed of several layers of hidden casual variables. Besides, there are connections exists between the layers and no connections between units within each layer [12]. It is the combination of probability and statistics with ML and neural networks. Figure 1 shows different types of DBN.

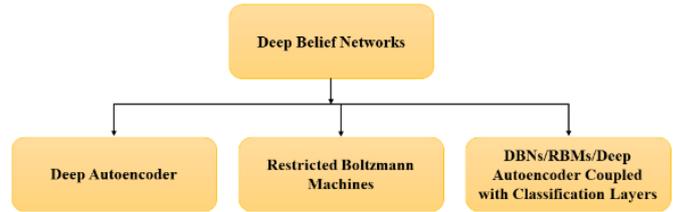

**Figure 1.** Classification of DBN

### B. Autoencoder

An unsupervised method is an autoencoder where the input is given as a vector. The network attempts to match and the output is the same as the input vector. One can generate a lower or higher dimensionality illustration of the data by getting the input and varying the recreating the input with its dimensionality. Data encoding operation (i.e., feature compression) is executed in the network with a small dimension of hidden layers. A denoising autoencoder can play an important role in order to eliminate the noise and reconstruct the original input from the noisy input. Figure 2 illustrates a basic autoencoder.

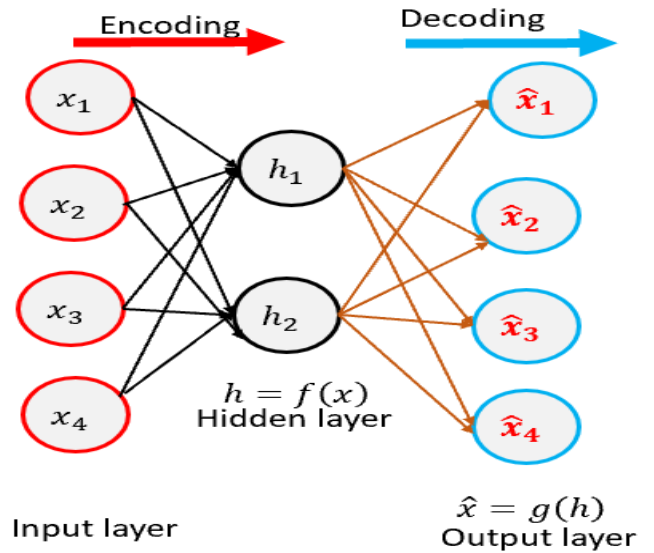

**Figure 2.** Autoencoder

### C. Recurrent Neural Network

A recurrent neural network (RNN), a subset of neural networks, which is connected between nodes and form a directed graph as shown in Figure 3. This makes the network in its internal state. It permits to show dynamic sequential behavior. They use their internal memory to process arbitrary sequences of input and the signal travels both forward and backward by creating loops in the network [13-15].

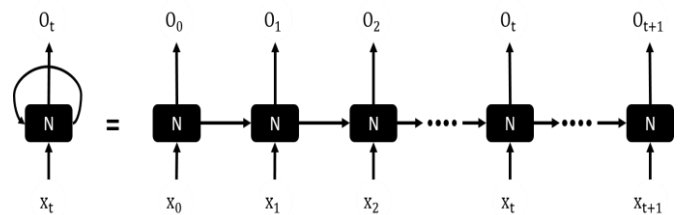

**Figure 3.** Recurrent Neural Network

Typically, it is more complex to train RNNs due to the disappearance of the gradients. However, the improvements in architecture and training have formed various RNNs. This model is simpler to train. The long short-term memory (LSTM), an improved system of RNN, was first brought by Hochreiter and Schmidhuber in 1997 [16]. LSTM is making a major change in speech recognition and set a revolutionary record on some traditional models in certain speech applications. It is introduced to solve RNNs short term memory problem. LSTM units connect to the situation in the following time stage. The configuration of the units that accumulate information is called a memory cell.

### D. Convolutional Neural Network

Convolutional neural network (CNN) is a portion of deep NN that processes as well as analyze visual imagery input. If a colored or grayscale image is considered as input, then the image will be stored in pixels like 2D array. In addition, CNNs are also applied for managing audio spectrograms with 2D arrays. However, the model of CNN contains three kinds of layers, including classification layers, pooling layers and convolution layers [15, 17]. An illustration of CNN is shown in Figure 4.

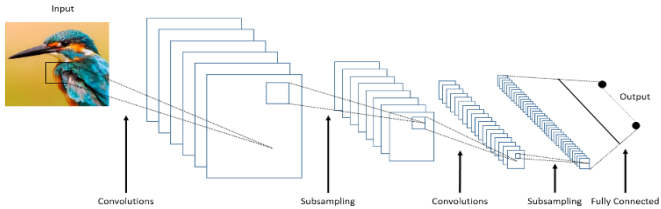

**Figure 4.** Convolutional Neural Network

### E. Generative Adversarial Networks

GANs are deployed in unsupervised ML, where 2 neural networks contest against one another in a game of zero-sum to overcome one another. It is introduced by the work of Goodfellow. Figure 5 shows the block diagram of GAN. The generator produces output data using the similar features as real time data by using input data. Then, the discriminator analyze the real data, whether the input is real or fake [18]. There is a wide range of applications in GAN system, including optical flow estimation [98], caption generation [97], image enhancement [96], and DCGAN for Facebook [99].

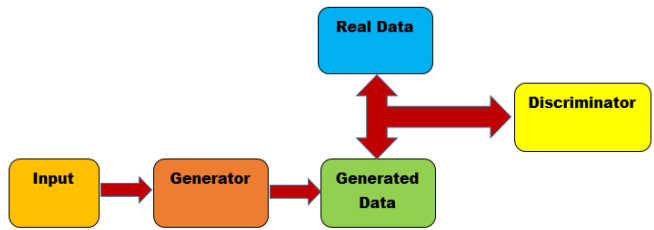

**Figure 5.** Generative Adversarial Networks

### F. Recursive Neural Network

Recursive neural networks relate a number of weights recursively. It has a number of inputs. At first, the primary 2 inputs are nurtured in the model as one. A node output is then considered as an input for the following node. Many natural language processing and image segmentation use this type of model.

## IV. COMPARISON BETWEEN SHALLOW LEARNING AND DL

This section provides a brief comparison between DL and shallow learning algorithms. DL has multiple layers, as shown in Figure 6. Besides, in DL, a deep network has several hidden layers, while shallow neural networks typically have 1-hidden layer. The neuron layers are linked with adaptive weights, besides the neighbor network layers are generally staying associated. However, there are two kinds of shallow network architecture: supervised and unsupervised. In supervised learning, the labels remain known to learn a work. Moreover, feature extraction is achieved individually.

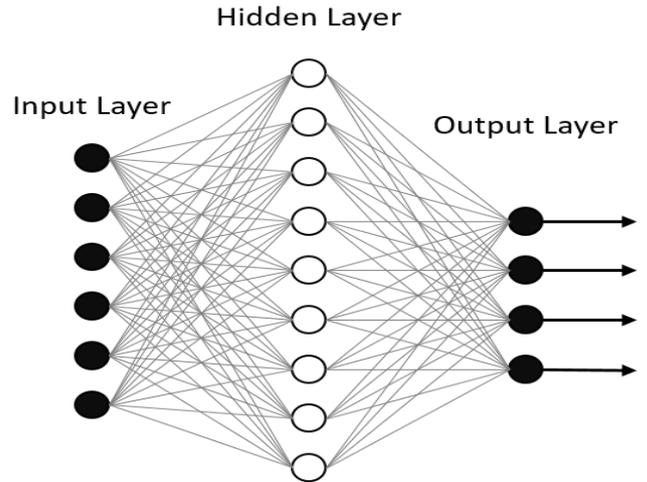

**Figure 6.** Shallow Neural Network

This forms of DL model derives higher-level features from the raw input with the help of its multiple hidden layers. Figure 7 illustrates a deep neural network. There are several levels between the input layer and output layer; the output layer is considered as higher level, and input layer are considered as lower level. From the lower-level concepts, higher-level concepts are defined. Although feature extraction can be obtained from the few initial layers of DL network. The DL architecture is of three types: unsupervised, hybrid and supervised. Advance feature extraction in shallow neural networks is performed separately because they have only one hidden layer. However, deep networks are capable of learning. However, with great computational power, several GPUs are needed for DL methods and it costs too much time to train DL models [19].

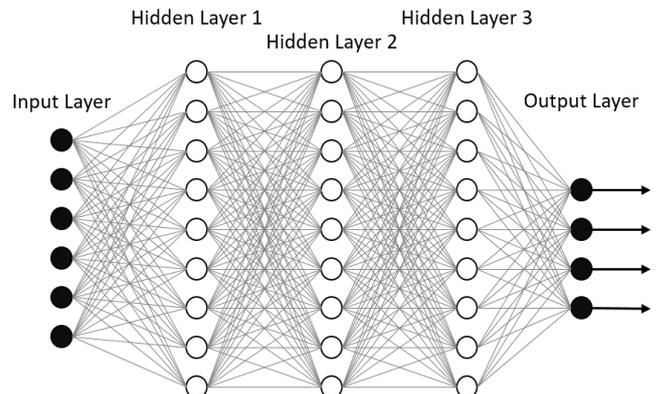

**Figure 7.** Deep Neural Network

| Cyber Attack | DL Method | Dataset | Research Paper |
|---|---|---|---|
| Malware Detection and Classification | CNN | Microsoft Malware Classification Challenge [108] | [41] |
| | Autoencoder | Comodo Cloud Security Center [119] | [33] |
| | Autoencoder | Dataset of call sequence in Public malware API [130] | [42] |
| | CNN | Genome Project [120], McAfee Labs | [32] |
| | CNN RNN | Maltrieve Private, Virus Share [121] | [24] |
| | CNN RNN | Unknown | [25] |
| | CNN (Dynamic) | Unknown | [39] |
| | DNN | Private data of Jotti commercial | [29] |
| | DNN | Internal Microsoft dataset | [40] |
| | DNN | DREBIN [107] | [45] |
| | DNN | Microsoft corporation provided dataset | [44] |
| | Autoencoders (Denoising) | C4 Security dataset | [43] |
| | RNN | Internal Microsoft dataset | [23] |
| | RNN | Virus Total [117], Alexa [60] | [36] |
| | RBM | Contagio [118] Google Play Store [116] | [21] |
| | RBM | Contagio [118], Google Play Store [116], Genome Project [120] | [22] |
| | RBM | Comodo Cloud [119], Security Center | [26] |
| | RBM | Virus share [121],Google play store [116] | [35] |
| | RBM | Self-generated dataset | [34] |
| | RBM | VirusTotal [117], DREBIN [107], Google Play [116], Genome Project [120] | [27] |
| | RBM | Comodo Cloud [119], Security Center | [28] |
| | RBM | Unknown | [38] |
| | RBM | Unknown | [31] |
| | Autoencoder | Challenge of Classification of Microsoft Malware[108],NSL-KDD [115] | [46] |
| | DNN | Malware of traffic data, VirusShare, Kaspersky, MalShare, Malware sample | [37] |

*Table 1.* DL methods for Malware Detection and Classification

| Cyber Attack | DL Method | Dataset | Research Paper |
|---|---|---|---|
| Intrusion Detection | Autoencoder | KDDCUP 1999 [114] | [56] |
| | Autoencoder | KDDCUP 1999 [114] | [62] |
| | Autoencoder | KDDCUP 1999 [114] | [63] |
| | Autoencoder | NSL-KDD [115] | [61] |
| | Autoencoder | Network Experiments and Open-Car-Test-bed | [60] |
| | Autoencoder | NSL-KDD [115] | [70] |
| | Autoencoder | CIDDS-001 [127] | [54] |
| | Ladder Networks (Autoencoder) | KDD 1999 [114] | [71] |
| | Autoencoder RBM | KDD 1999 [114] | [53] |
| | Autoencoder | Custom | [55] |
| | CNN | CTU-13 [124], IXIA [131] | [56] |
| | CNN (dilated) Autoencoder | CTU-UNB [124, 125], Contagio-CTU-UNB [125] | [20] |
| | DNN | KDDCUP 1999 [114] | [58] |
| | DNN | simulator of Cooja network | [59] |

| | | |
|---|---|---|
| DNN | NSL-KDD [115] | [69] |
| RBM | KDDCUP 1999 [114] | [49] |
| RNN | KDDCUP 1999 [114] | [67] |
| RNN | UNM, ADFA-LD, KDDCUP-1998 [114] data sets | [65] |
| RNN | KDDCUP-1999 and additional [114], unique data | [66] |
| RNN | Custom | [46] |
| RNN | KDD 1999 [114] | [68] |
| RNN | KDD 1999 [114] | [64] |
| RNN | NSL-KDD [115] | [57] |
| RBM | NSL-KDD [115] | [47] |
| RBM | KDD 1999 [114] | [50] |
| RBM | NSL-KDD [115] | [51] |
| RBM | NSL-KDD [115] KDDCUP 1999 [114] UNSW-NB15 | [48] |
| RBM Autoencoder | KDD 1999[114] | [52] |
| RBM RNN | CTU-13[124] | [158] |
| Autoencoder | Challenge of Microsoft Classification of Malware [108], NSL-KDD [115] | [46] |

*Table 2.* DL methods for Intrusion Detection

| Cyber Attack | DL Method | Dataset | Research Paper |
|---|---|---|---|
| File Type Identification | Autoencoder | Internal Dataset | [81] |
| Identification of Network | Autoencoder | Honeypot dataset resulted internally | [83] |
| Traffic | Autoencoder | ISCX-VPN-nonVPN-traffic dataset[164] | [82] |
| Spam Identification | Autoencoder | EnronSpam [126], PU1, PU2, PU3, PU4 | [84] |
| | RBM | EnronSpam [126] SpamAssassin [127] LingSpam [128] | [85] |
| Impersonation Attacks | Autoencoder | AWID[156] | [86] |
| User Authentication | Autoencoder | Custom | [87] |
| DGA | CNN | Alexa [104], Private Dataset | [88] |
| | GAN | Alexa [104] | [89] |
| | RNN | Alexa [104], OSINT [105] | [90] |
| | RNN | Alexa [104], DGArchive [106], OSINT [105] | [91] |
| | CNN RNN | Alexa [104], OSINT [105] | [92] |
| | CNN RNN | Alexa [104], DGArchive [106] | [93] |
| | RNN | Alexa [104], OSINT [105] | [94] |
| | RNN | Malware Capture Facility Project Dataset | [95] |
| Attack of Drive-by | CNN | KDD 1999 [114] | [119] |
| Download | CNN | honeypot setup, Malware Bytes,Malware domain list, Alexa [104] | [118] |
| Traffic Identification | CNN | ISCX VPN-nonVPN traffic dataset [164] | [163] |
| Insider Threat | DNN RNN | CERT Dataset v6.2 [123] | [166] |
| Anomaly Detection | RNN | Custom | [169] |
| Keystroke Verification | RNN | Custom | [170] |
| False Data Ejection | RBM | Custom | [172] |

*Table 3.* DL methods for Other Frequent Cyber Attack

| Dataset Name | Features | Reference |
|---|---|---|
| KDD Cup 99 | Intrusion Detection, R2L, DoS, Probing | [114] |
| NSL-KDD | Network Intrusion Detection | [115] |
| CTU-13 | Scenarios of thirteen captures of various samples in botnet are included in this dataset. Every scenario is taken in a file like pcap. That file consists of every packets of 3 kinds of traffic. | [59] |
| Alexa | Alexa offers us access to a set of web sites. Alexa can be expected legitimate. | [104] |
| AWID | Comprehensive WiFi network benchmark dataset. Detect impersonation attacks | [100] |
| DGA | 38 classes with 168,900 samples | [106] |
| CTU | Consists of different, real botnet attacks and labels. | [103] |
| OSINT | OSINT DGA feeds from Bambenek Consulting. DGA (Domain Generation Algorithm) are used by malware | [105] |
| VirusShare | A repository of 38,005,488 malware samples. | [121] |
| DREBIN | 1,20,000 android applications are contained in this dataset. | [107] |
| Microsoft Malware Classification Challenge | Every malware has a unique value of 20-character hash recognizing the file and a unique Id, and a Class. An integer value is represented names of nine-family. This malware can belong. | [108] |
| CERT Insider Threat Dataset v6.2 | System logs spanning 516 days are included. This dataset contains over 130 million events. Approximately 400 of them are malicious. | [109, 110] |
| EnronSpam | The total number of emails (legitimate and spam) is 5975. The ratio of spam to legitimate rate is 1:3. | [111] |
| SpamAssassin | It can differentiate effectively between non-spam and spam in between cases of 95% to 100%, relying on the kind of mail taken and Bayesian filter with training. | [112] |
| Malware Training Sets | Zeus:2014 samples, Crypto:2024 samples, Locker:434 samples, APT1:292 samples. | [123] |
| CIDDS-001 | The dataset consists a large number of traffic instances. 153026 instances are collected from External Server. 172839 instances are collected from OpenStack Server. | [127] |
| Public malware API call sequence dataset | API call sequences (Malware): 42,797, API call sequences (goodware): 1,079. Each API call sequence: First 100 API calls that are consecutively non-repeated | [130] |
| ISCX VPN-nonVPN traffic dataset | It has a categories of 14-traffic: P2P, VOIP, VPN-P2P,VPN-VOIP, etc. Wireshark and tcpdump were used for capturing the traffic generated about 28GB of data. | [131] |

*Table 4.* Explanation of Cyber Security Datasets in DL

|  | Malicious | Benign |
|---|---|---|
| Malicious | True Positive (TP) | False Negative (FN) |
| Benign | False Positive (FP) | True Negative (TN) |

Predicted level

| Confusion Matrix | |
|---|---|
| True positive rate, TPR /recall/hit rate | $\dfrac{TP}{TP+FN}$ |
| True negative rate, TNR or specificity | $\dfrac{TN}{TN+FP}$ |
| False alarm rate (FAR) or False positive rate, FPR | $\dfrac{FP}{TN+FP}$ |
| False negative rate, FNR or Miss rate | $\dfrac{FN}{TP+FN}$ |
| Precision | $\dfrac{TP}{TP+FP}$ |
| Negative predicted value | $\dfrac{TN}{TN+FN}$ |
| Accuracy | $\dfrac{TP+TN}{TP+TN+FP+FN}$ |
| F1 Score | $\dfrac{2\times Precision\times Recall}{Precision+Recall}$ |

**Figure 8.** Several performance metrics

However, DL takes too much time to analyze and extract relevant information from the huge amount of data and the data is not formed properly.

Table 1 summarizes various DL methods applied by researchers for malware detection and classification. Most researchers use restricted Boltzmann machine (RBM) method. Table 2 summarizes various DL methods applied for intrusion detection. Most researchers use autoencoder and RNN method. Table 3 summarizes the DL method used in order to detect other type of cyber-attacks.

KDD Cup 99 dataset formed for the challenge of KDD in 1999 is one of the most commonly used datasets in order to detect the various type of intrusions. KDD means Knowledge Discovery. About more than four million network traffic records exist in this dataset. Twenty two different types of attacks are contained in this dataset that can be categorized into four families such as denial-of-service (DoS), R2L, for example, predicting the password, U2R, and probing. The other datasets used in various research papers for the classification of various threats have been described in Table 4 with short details. Several performance metrics are depicted in Figure 8.

## V. PERFORMANCE ANALYSIS OF DL METHODS APPLIED IN VARIOUS RESEARCH PAPERS

DL models have shown significant improvements over traditional ML-based solutions, signature-based methods and rule-based methods in order to address cybersecurity problems. Table 5 illustrates the performance results achieved adopting different DL models. The results are reported in terms of precision, false negative rate (FNR), classification accuracy, F1-score, true positive rate (TPR), etc. We have reviewed 85 papers. From the review, it can be seen that most researchers have focused on malware classification and detection of various types of intrusion in the network. Cyber-physical autonomous systems which is not only sensor-based but also communication-enabled (e.g., automotive systems), biometrics behavioral (i.e., signature dynamics) are considered as increasing areas for DL applications of security.

As we become more reliant on network-connected devices, we will see an increase in the number of cyber-physical systems and computational systems, each having its own set of attack vectors owing to its unique baseline. For malware and intrusion detection, RBMs were the most often utilized DL technique. RNNs were another popular solution for tackling the largest range of cyber security challenges feasible (i.e., network intrusions, cyber-physical intrusions, malware, host intrusions and names of malicious domain).

The large use of RBMs and autoencoders, around 50%, is most likely owing to a scarcity of labeled data, and unlabeled data is pre-trained and fine-tuned using a little quantity of labeled data. RNNs are likely popular because many cyber security jobs or data may be treated as a time series problem. This is beneficial to RNNs.

Conclusions on the success of any approach are difficult to make since various studies utilize various datasets and measurements. Certain tendencies, however, are remarkable. The performance of various areas of the security business varied greatly. Domains constructed employing a variety of techniques seem to have the most consistent DGA-produced hazardous domains, with TPRs ranging from 1% to 1.5 % and accuracy values ranging from 0.9959 to 0.9969, equivalent to 96.01 to 99.86%. Network intrusion detection techniques, on the other hand, have a performance range of 92.33 to 100 percent with a TPR of 1.58 to 2.3 percent and an accuracy range of 44 to 99 percent. A high classification accuracy of 99.72% is reported for RBM when applied to a custom dataset [34], while accuracy of 99.80% is achieved by LSTM for KDD Cup 99 dataset [66]. Historically, the capacity to detect network intrusions has significantly been reliant on the kind and quantity of attacks carried out. Another crucial component influencing overall performance was the training set's relationship between benign and dangerous data. This quandary stems from the difficulties of getting legally harmful materials. Because authentic data might be difficult to get, data is often generated using viral simulations and reverse engineering.

| Methods | Data used | Paper | Precision | FNR | Accuracy | F1-Score | TPR |
|---------|-----------|-------|-----------|-----|----------|----------|-----|
| DBN | KDD CUP 99 | [47] | 92.33% | | 93.49% | | |
| LR-DBN | 10% KDDCUP'99 | [49] | 97.9% | 2.47% | | | |
| DBN | Netflow | [31] | | | 96.7% | | |
| DL RBM | NSL-KDD, KDD CUP 1999, UNSW-NB15 | [48] | 81.95%, 94.43%, 83.40% | | 90.99%, 97.11%, 95.84% | | 77.48%, 92.77%, 79.19% |
| RBM | [118], [120], [116] | [22] | 95.77% | | 96.76% | | 97.84% |
| DBN | 40% NSL-KDD | [50] | | | 97.5% | | |
| RNN | NSL-KDD | [57] | | | 83.28% | | |
| Autoencoder | KDD CUP 99 | [71] | | | 99.18% | | |
| DBN based PNN | KDD CUP 99 | [80] | 93.25% | 0.62% | 99.14% | | |
| CNN and RNN (LSTM and Softmax layer) | Virus Share [121], Maltrieve Private | [24] | 85.6% | | 89.4% | | 89.4% |
| DBN | | [31] | | | 96.7% | | |
| CNN | Genome Project, McAfee Labs (malware samples: 2475 and benign samples: 3627) (benign: 9268 and malware: 9902) | [32] | 99.0%, 27% | | 98.0%, 80% | 97.0%, 78% | 95.0%, 85% |
| Autoencoder | Comodo Cloud Security Center | [33] | | | 95.64% | | |
| RBM | - | [26] | | | 96.6% | | |
| RBM | Custom | [34] | | | 99.72% | | 90.1% |
| RBM | EnronSpam | [85] | | | 93.4% | | |
| RSTNN | Custom | [36] | 97.6% | | | 96.9% | 96.2% |
| DNN | - | [37] | 97.1% | | | | 100% |
| Recurrent SVM | Alexa, OSINT | [92] | 92.06% | | 99.69% | 92.60% | 93.14% |
| Bidirectional LSTM | Alexa, OSINT | [92] | 92.32% | | 99.64% | 92.70% | 93.09% |
| Fusion of CNN and LSTM | Alexa, OSINT | [92] | 91.59% | | 99.59% | 92.06% | 92.53% |
| CNN | Alexa, DGArchive | [93] | | | 99.18% | | 72.89% |
| LSTM | Alexa, DGArchive | [93] | | | 98.96% | | 74.05% |
| DNN | | [99] | | | 97% | | |
| CNN | Malimg Dataset, Microsoft Malware Dataset | [41] | | | 98.52%, 98.99%, 99.57% | | |
| DNN | DREBIN | [45] | | 6.37%, 3.96% | 95.93%, 98.35% | | |
| RNN | UNM, ADFA-LD, KDDCUP 1998 [114] data sets | [65] | 99.31% | 4.62% | 96% | 97.31% | 95.38% |
| RNN | KDD Cup 99 | [67] | | | 96.93% | - | 98.88% |
| RNN | KDD Cup 99 | [68] | 84.6% | | 77.55% | - | 73% |
| LSTM | 10% KDD Cup 99 | [64] | - | | 93.85% | - | - |
| LSTM | 10% KDD Cup 99 | [72] | 98.8% | | 96.93% | - | - |
| LSTM | KDD Cup 99 | [66] | - | | 99.8% | - | - |
| LSTM | KDD Cup 99 | [73] | 98.95% | | 97.54% | - | - |
| GRU | Netflow | [74] | - | | 84.15% | - | - |
| CNN | CTU-UNB datasets | [20] | 98.44% | | - | - | 98% |
| CNN | Netflow | [75] | 93% | | - | - | 92% |
| CNN | Netflow | [76] | - | | 92% | | - |
| 1D-CNN | ISCX dataset | [77] | 97.3% | | - | - | 96% |
| CNN | Netflow | [78] | - | | 99.41% | | - |
| Stacked Autoencoder | AWID | [101] | 86.15% | | 92.18% | | |
| DBN, RNN | CTU dataset | [102] | 81.26%, 68.63% | | | | 99.34%, 70.35% |
| CNN | Microsoft Malware Classification Challenge (2015) | [129] | | | 99.24% | | 100% |

*Table 5.* Performance Analysis of various DL methods

When employing DL-driven security technology, some difficulties may arise. The model's accuracy can be viewed as a significant impediment.

The use of any new tool, especially DL tools, is universally frowned upon because they are ultimately black boxes. As a result, when errors occur, determining the cause is impossible, and unlike DL applications such as the marketing sector, larger costs and hazards are associated with cybersecurity missteps. A cybersecurity analyst may waste time analyzing false alarms, or an automated response to intrusion detection may erroneously restrict access to critical services. Furthermore, a DL tool can completely ignore a cyber-attack. Another barrier to adoption is that many of the currently available systems focus on a specific hazard, such as virus detection. Researchers should investigate methods for generalizing or combining multiple DL approaches in order to cover a broader range of attack vectors and provide a more comprehensive solution. Multiple DL detection techniques must be used concurrently, and information gathered by various techniques may also be used to improve local performance.

Cybersecurity has become an important issue for IoT since IoT can contribute to managing pandemics, particularly the novel coronavirus disease (COVID-19). One example of the use of IoT for COVID-19 is to mitigate the causative virus from being spreading. This can be done by the screening of temperature, tracing the contacts, and several other ways. Detecting early cases of the infection, tracing, and then isolating the suspected patients can be done with IoT. Note that IoT-driven healthcare systems and IoT-driven COVID-19 diagnosis systems are emerging techniques that can be useful to patients and doctors. Another example is facilitating the new lifestyle during COVID-19, including home-office, distant learning, fitness training at home, etc. These activities enable the running of businesses, educational institutions, government offices without risking the people's health. Another use case of IoT is to resolve machinery issues for controlling medical inventory, tracking tagged nebulizers, oxygen cylinders, and other medical equipment. For tackling a pandemic, IoT can be used along with other techniques such as near field communication, radio frequency identification, WiFi, light fidelity, sensor networks, etc. These technologies require small portable devices that have low computation power and low battery life. As a result, ensuring cybersecurity for small IoT devices is a more challenging task compared to traditional computers, server, smartphones and laptops. Cyber attacks evolve rapidly, so it is difficult to incorporate security measures in IoT devices quickly. Unless the cyber attacks are mitigated, IoT cannot be effectively used in controlling pandemics. Security threats such as phishing, spamming, ransomware, Distributed DoS [137-143] may affect the reliability of IoT-driven healthcare and COVID-19 diagnosis [132-136] systems. Hence, understanding the possible security threats and finding appropriate mitigation techniques is essential in the context of IoT and other networking scenarios.

## VI. CONCLUSIONS

This paper focuses on the use of DL in improving the security system. As attacks of malicious against cyber system networks are advancing, the cyber defender needs to be more advanced. Cybersecurity personnel should have the capability to remark and employ original signatures to identify original attacks. DL approaches to cybersecurity applications offer a smart opportunity to identify novel malware variants and attacks of zero-day. In this review, we have described the applications of DL systems to different types of cybersecurity attack types. These attacks are mainly application software, targeted networks, data and host system. Likewise, this paper illustrates that the standard datasets are very important to advancing DL in the cybersecurity domain. The paper aims to draw a complete review of DL methods, the needs of DL in cybersecurity, and to encourage future research of DL in cybersecurity. Finally, this article discusses the use case scenarios of IoT in the context of COVID-19, and highlights the importance of cybersecurity for IoT devices.

## Author Biographies

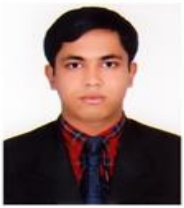

**Prajoy Podder** worked as a Lecturer in the Department of Electrical and Electronic Engineering in Ranada Prasad Shaha University, Narayanganj-1400, Bangladesh. He completed B.Sc. (Engg.) degree in Electronics and Communication Engineering from Khulna University of Engineering & Technology, Khulna-9203, Bangladesh. He is currently pursuing M.Sc. (Engg.) degree in Institute of Information and Communication Technology from Bangladesh University of Engineering and Technology, Dhaka-1000, Bangladesh. He is a researcher in the Institute of Information and Communication Technology, Bangladesh University of Engineering & Technology, Dhaka-1000, Bangladesh. He is regular reviewer of Data in Brief, Elsevier and Frontiers of Information Technology and Electronic Engineering, Springer, ARRAY, Elsevier. He is the lead guest editor of Special Issue on Development of Advanced Wireless Communications, Networks and Sensors in American Journal of Networks and Communications. His research interest includes machine learning, pattern recognition, neural networks, computer networking, distributed sensor networks, parallel and distributed computing, VLSI system design, image processing, embedded system design, data analytics. He published several IEEE conference papers, journals and Springer Book Chapters.

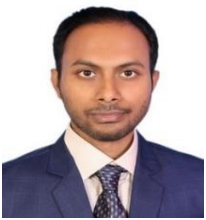

**Subrato Bharati** received his B.S. degree in Electrical and Electronic Engineering from Ranada Prasad Shaha University, Narayanganj-1400, Bangladesh. He is currently working as a research assistant in the Institute of Information and Communication Technology(IICT), Bangladesh University of Engineering and Technology, Dhaka, Bangladesh. He is a regular reviewer of several Wiley, Elsevier and Springer International Journals. He is an associate editor of Journal of the International Academy for Case Studies. He is a member of scientific and technical program committee in some conferences including CECNet 2021, ICONCS, ICCRDA-2020, ICICCR 2021, etc. His research interest includes bioinformatics, medical image processing, pattern recognition, deep learning, wireless communications, data analytics, machine learning, neural networks, and feature selection. He published several journals paper, and also published several IEEE, Springer reputed conference papers. He published Springer and Elsevier, De Gruyter, CRC Press and Wiley Book chapters as well.

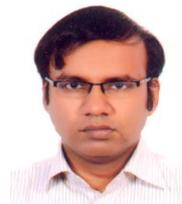

**M. Rubaiyat Hossain Mondal**, PhD is currently working as a faculty member in the Institute of Information and Communication Technology (IICT) at Bangladesh University of Engineering and Technology (BUET), Bangladesh. He received his Bachelor's degree and Master's degree in Electrical and Electronic Engineering from BUET. He joined IICT, BUET as a faculty member in 2005. From 2010 to 2014 he was with the Department of Electrical and Computer Systems Engineering (ECSE) of Monash University, Australia from where he obtained his PhD in 2014. He has authored a number of articles in reputed journals, conferences and book chapters. He is an active reviewer of several journals published by IEEE, Elsevier and Springer. He was a member of the Technical Committee of different IEEE International conferences. His research interest includes artificial intelligence, bioinformatics, image processing, wireless communication and optical wireless communication.

**Pinto Kumar Paul** received his B.Sc degree in CSE from Daffodil International University, Dhaka, Bangladesh. He currently working as a Lecturer in the Department of CSE, Ranada Prasad Shaha University, Narayanganj-1400, Bangladesh. His research interest includes NLP, Image Processing and Internet of Things.

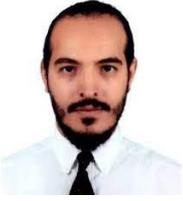 **Utku Kose**, PhD received the B.S. degree in 2008 from computer education of Gazi University, Turkey as a faculty valedictorian. He received M.S. degree in 2010 from Afyon Kocatepe University, Turkey in the field of computer and D.S. / Ph. D. degree in 2017 from Selcuk University, Turkey in the field of computer engineering. Between 2009 and 2011, he has worked as a Research Assistant in Afyon Kocatepe University. Following, he has also worked as a Lecturer and Vocational School - Vice Director in Afyon Kocatepe University between 2011 and 2012, as a Lecturer and Research Center Director in Usak University between 2012 and 2017, and as an Assistant Professor in Suleyman Demirel University between 2017 and 2019. Currently, he is an Associate Professor in Suleyman Demirel University, Turkey. He has more than 100 publications including articles, authored and edited books, proceedings, and reports. He is also in editorial boards of many scientific journals and serves as one of the editors of the Biomedical and Robotics Healthcare book series by CRC Press. His research interest includes artificial intelligence, machine ethics, artificial intelligence safety, optimization, the chaos theory, distance education, e-learning, computer education, and computer science.